\documentclass[preprint2]{aastex6}
\usepackage{graphicx}
\usepackage{amsmath}	
\usepackage{amssymb} 

\newcommand{\HI}{H\,{\sc i}}
\newcommand{\HeI}{He\,{\sc i}}
\newcommand{\LiI}{Li\,{\sc i}}

\newcommand{\OI}{[O\,{\sc i}]}
\newcommand{\CaI}{Ca\,{\sc i}}
\newcommand{\CaII}{Ca\,{\sc ii}}

\newcommand{\kms}{\,km\,s$^{-1}$}

\begin{document}

\title{The curious case of PDS 11: a nearby, $>$10 Myr old, classical T Tauri binary system}
\shorttitle{PDS 11: a nearby, $>$10 Myr old CTTS binary system}

\author{Blesson Mathew\altaffilmark{1} and P. Manoj}
\affil{Department of Astronomy and Astrophysics\\
Tata Institute of Fundamental Research \\
Homi Bhabha Road, Colaba, Mumbai 400005, India}

\author{B. C. Bhatt}
\affil{Indian Institute of Astrophysics\\
Koramangala, Bangalore 560034, India}

\author{D. K. Sahu}
\affil{Indian Institute of Astrophysics\\
Koramangala, Bangalore 560034, India}

\author{G. Maheswar}
\affil{Aryabhatta Research Institute of Observational Sciences (ARIES)\\
Nainital 263002, India}

\and

\author{S. Muneer}
\affil{Indian Institute of Astrophysics \\
Koramangala, Bangalore 560034, India}

\shortauthors{B. Mathew et al.}
  
\altaffiltext{1}{blesson.mathew@tifr.res.in}

\begin{abstract}
  We present results of our study of PDS 11 binary system, which belongs to a rare class of
  isolated, high galactic latitude T Tauri stars.
  Our spectroscopic analysis reveals that PDS 11 is a M2$-$M2 binary system with both components showing
  similar H$\alpha$ emission strength. Both the components appear to be accreting, and
  are classical T Tauri stars. The lithium doublet \LiI~$\lambda$6708, a signature of youth,
  is present in the spectrum of PDS 11A, but not in PDS 11B. From the application of lithium
  depletion boundary age-dating method and a comparison with the \LiI~$\lambda$6708 equivalent width
  distribution of moving groups, we estimated an age of 10$-$15 Myr for PDS 11A.
  Comparison with pre-main sequence evolutionary models indicates that PDS 11A is
  a 0.4 M$_\odot$ T Tauri star at a distance of 114$-$131 pc.
  PDS 11 system does not appear to be associated with any known star
  forming regions or moving groups. PDS 11 is a new addition,
  after TWA 30 and LDS 5606, to the interesting class of old,
  dusty, wide binary classical T Tauri systems in which both components are actively accreting.
\end{abstract}


\section{Introduction} \label{sec:intro}

T Tauri stars are low-mass (K \& M spectral types) young stars which are in their pre-main
sequence phase of evolution \citep[e.g.,][]{Joy45,Bertout89,Herczeg14}.
They are often associated with cloud complexes such as
Taurus, Orion and Ophiuchus \citep{Herbig62,Herczeg14,Appenzeller89}.
However, T Tauri stars are also found in isolated regions above the Galactic
plane and far from any dark clouds \citep{Reza89,McGehee13,Elliott16}. 
TW Hydrae was the first such `isolated T Tauri star' identified, which was later
found to be part of an association of
about three dozen members, known as TW Hydrae association \citep[TWA,][]{Kastner97,Zuckerman04,Mamajek16}.
They were not runaway stars from molecular clouds but
formed in situ in the present region $\sim$10 Myr ago, which is now
devoid of molecular gas \citep{Rucinski83,Tachihara09}.
Since then several such nearby, young associations have been identified such as, the $\beta$
Pictoris moving group, the AB Doradus moving group, the Tucana/Horologium
association etc., within 100 pc of the Sun \citep[see][for a review]{Zuckerman04,Torres08,Mamajek16}.

%figure 1
\begin{figure}
\figurenum{1}  
\plotone{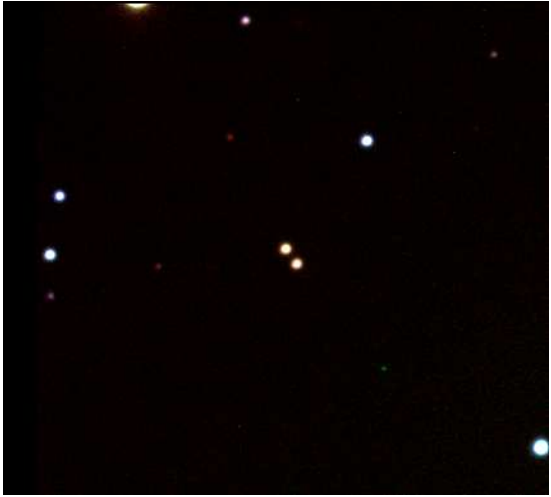}
\caption{$BVR$ color composite of PDS 11, with North up and East to the left.
  This 4$\arcmin$$\times$4$\arcmin$ composite image is constructed
  from our observations with the HCT and is color coded as blue, green
  and red in $B$, $V$, $R$, respectively.
  The image is centered on PDS 11 binary system,
  with the north-eastern component being PDS 11A.}
\end{figure}

In this paper we present a  detailed analysis of the high galactic latitude (b = $-$30$^{\circ}$)
binary T Tauri star system  PDS 11  (GSC 04744-01367,  IRAS  04451-0539).
\citet{GregorioHetem92} have reported this to be a binary system
with components PDS (PDS stands for Pico dos Dias Survey) 11A and PDS 11B.
PDS 11A is the North-Eastern component of the binary (Figure 1).
PDS 11A and PDS 11B are present in the Washington double star (WDS) catalog,
with the component magnitudes being 14.76 and 15.34, respectively.
The position angle (PA) and separation were found to be 216$^{\circ}$ and
8.8$\arcsec$, respectively \citep{Mason01}. Our analysis indicate that PDS 11 is a
young (10$-$15 Myr), nearby (114$-$131 pc) binary T Tauri
system, where both the components are possibly accreting. The paper is arranged as follows. 
Section 2 describes the optical and near-IR observations. 
The results from this study are presented in Sect. 3. A detailed discussion of the key results is given
in Sect. 4, and finally in Sect. 5 we present our conclusions. 

\section{Observations and data reduction} 

\subsection{Optical spectroscopy}

Optical spectra of both the components of the PDS 11 system 
in the wavelength range 3800$-$9000 \AA~were obtained 
with the Himalayan Faint Object Spectrograph Camera (HFOSC)\footnote{Further details of the instruments and 
telescopes is available at http://www.iiap.res.in/iao/hfosc.html} 
mounted on the 2-m Himalayan Chandra Telescope (HCT). The wavelength
range was covered using Grism 7 (blue region, 3800$-$5500 \AA) and Grism 8 (red
region, 5500$-$9000 \AA). The grisms in combination with 1.92$\arcsec$ wide
and 11$\arcmin$ long slit provide an effective resolving power of 
$\sim$ 900 in blue region and $\sim$ 1050 in red region. 
A spectrophotometric standard, Feige 34, was observed on the
same night and is used for flux calibration.
The observations were carried out on 2016 February 13.
The FeNe, FeAr lamp spectra were taken after each on-source exposure for wavelength calibration. 
The spectra were reduced in a standard manner after bias subtraction and flat
field correction using the standard tasks in Image Reduction and Analysis
Facility (IRAF)\footnote{IRAF is distributed by the National Optical Astronomy
  Observatories, which are operated by the Association of Universities for
  Research in Astronomy, Inc., under cooperative agreement with the National
  Science Foundation}. Further, the extracted
spectra were wavelength calibrated and flux calibrated.
The target was again observed on March 15 to confirm the spectral features observed.
The log of observations is given in Table 1.

\subsection{Near-infrared spectroscopy}

We also obtained near-IR spectra of PDS 11 with TIFR near-infrared spectrometer and imager
\citep[TIRSPEC;][]{Ninan14} mounted on 2-m HCT. The spectra were taken in 
$Y$ (1.02$-$1.20 $\mu$m), $J$ (1.21$-$1.48 $\mu$m), $H$ (1.49$-$1.78 $\mu$m) and $K$
(2.04$-$2.35 $\mu$m) bands, in combination with Grism and L3 slit (1.97$\arcsec$
wide and 300$\arcsec$ long). The effective resolving power is around 1200. 
The program stars PDS 11A and PDS 11B are co-aligned in the slit and
care is taken while reducing the spectra to extract them separately. 
A telluric standard HIP34768 (A1V spectral type) at nearby airmass was also
observed. The observations were carried out in dithered mode.
The log of infrared spectroscopic observations is given in Table
1. Argon spectra were obtained after the object spectra for wavelength
calibration. The wavelength calibrated object spectrum is divided 
with the the telluric spectrum whose hydrogen absorption lines were removed. 
The resultant spectrum is multiplied with a blackbody spectrum of 9230 K, 
corresponding to A1V spectral type of the telluric standard \citep{Kenyon95}. 
The final spectrum is normalized with respect to
the band center in $Y$, $J$, $H$ and $K$ bands.    

%table 1
\begin{deluxetable*}{llcccccc}
\tablecaption{Journal of spectroscopic observations \label{tab:table1}}
\tablehead{
\colhead{Object} & \colhead{Date} & \twocolhead{Optical} & \multicolumn{4}{c}{Infrared}\\
\colhead{}       & \colhead{}     & \twocolhead{Exp.time (s)} & \multicolumn{4}{c}{Exp.time (s)}\\
\colhead{}       & \colhead{}     & \colhead{Gr7/167l} & \colhead{Gr8/167l} & \colhead{$Y$} & \colhead{$J$} & \colhead{$H$} & \colhead{$K$}
}
\colnumbers
\startdata
PDS 11A   & 2016 Feb. 13 & 1800 & 1800  & . & . & . & . \\
          & 2016 Feb. 14 &  .   &  .  & 1000 & 1000 & 1000 & 1000 \\
          & 2016 Mar. 15 & 1800 & 1800 & . & . & . & . \\
PDS 11B   & 2016 Feb. 13 & 1800 & 1800 & . & . & . & . \\
          & 2016 Feb. 14 &  .  &  . & 1000 & 1000 & 1000 & 900 \\
          & 2016 Mar. 15 & 900 & 900 & . & . & . & . \\
Feige 34 & 2016 Feb. 13  & 600  & 600  & . & . & . & . \\
HIP 34768 & 2016 Feb. 14 & .  & . & 400 & 400 & 240 & 160 \\ 
\enddata
\end{deluxetable*}

\subsection{Optical photometry}

We imaged the 10$\arcmin$$\times$10$\arcmin$ region centered on
PDS 11 in $BVR$ passbands \citep{Bessell90} 
on 2016 March 19 using HFOSC.
The data reduction was carried out using various packages available in IRAF. 
Aperture photometry was performed on the program star and nearby field stars. 
The $B$, $V$, $R$ magnitudes of the nearby field stars were 
obtained using available SDSS photometry, which were converted 
to Bessell system using the transformation relations given 
in Lupton (2005)\footnote{https://www.sdss3.org/dr10/algorithms/sdssUBVRITransform.php}.
The magnitude of the program stars were calibrated 
differentially with respect to the nearby field stars.

\section{Results}

\subsection{Spectral analysis: optical and near-IR}

The spectra of both PDS 11A (Figure 2) and PDS 11B (Figure 3) 
look very similar due to the presence of
Balmer emission lines, from H$\alpha$ all the way up to H8 (8$-$2), \CaII~H \& K emission lines and
TiO absorption bands, albeit with different line strengths.
\citet{GregorioHetem92} reported an H$\alpha$ equivalent width (EW) of $-$20 \AA~for PDS 11A and
$-$42 \AA~for PDS 11B. The equivalent width measured from our spectra are
$-$23~\AA~and $-$27 \AA, which is quite different from the earlier
measurements, particularly for PDS 11B.
The EW and full width at half maximum (FWHM) of the prominent spectral lines are
given in Table 2.

%figure 2
\begin{figure}
\figurenum{2}  
\plotone{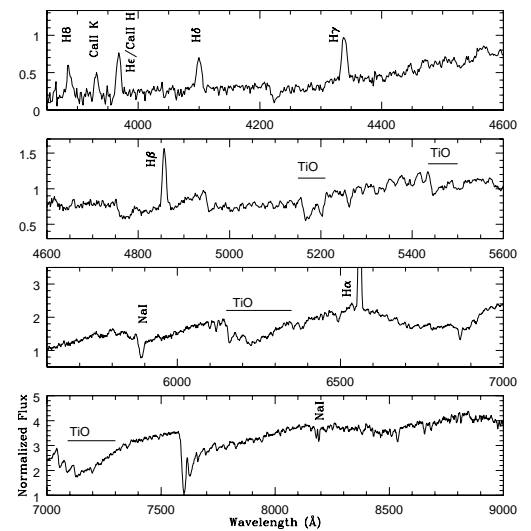}
\caption{Optical spectrum of PDS 11A. The spectrum is flux calibrated and is
  normalized at 5500 \AA. Prominent spectral lines are marked.}
\end{figure}

%figure 3
\begin{figure}
\figurenum{3}  
\plotone{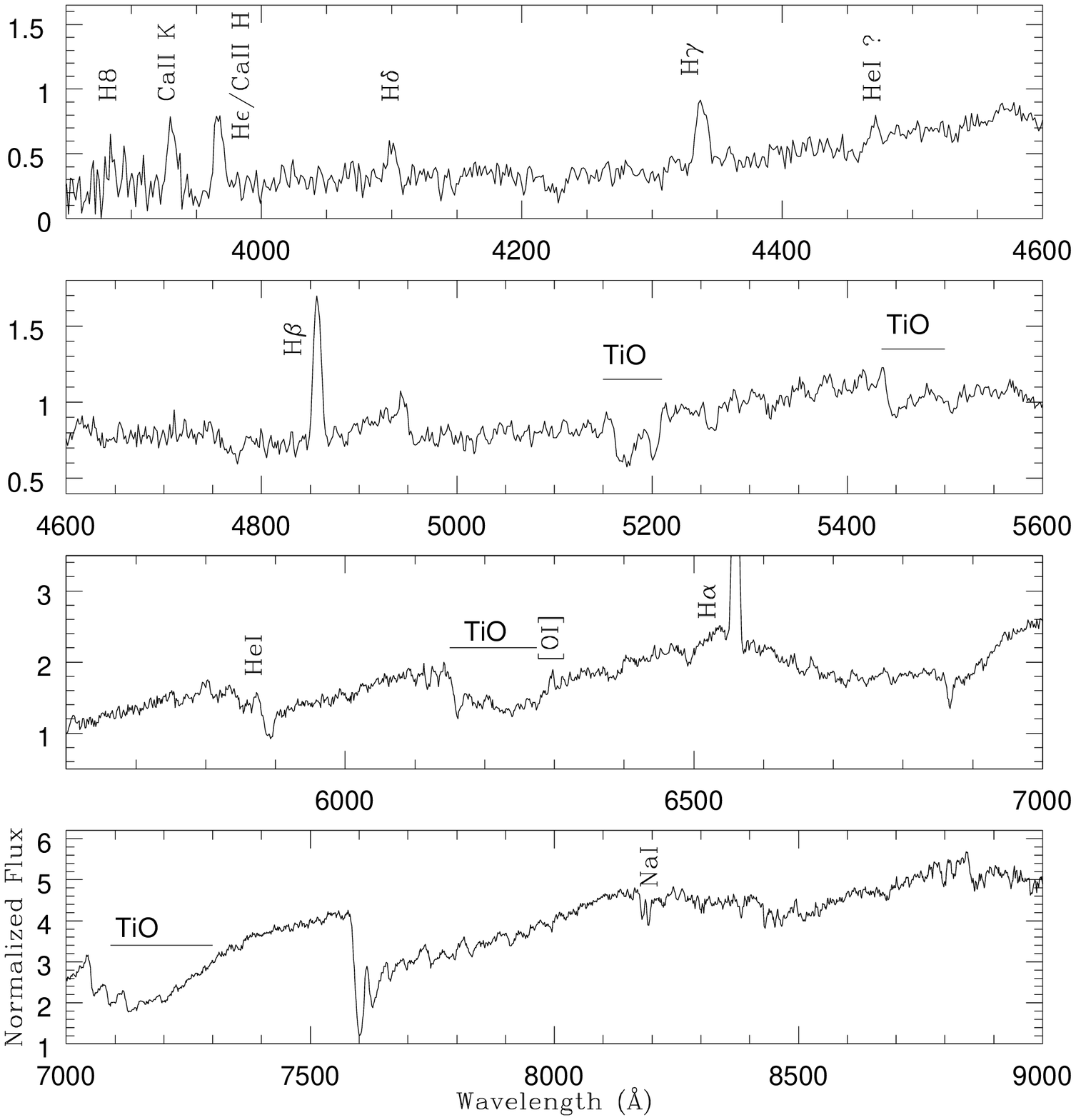}
\caption{Optical spectrum of PDS 11B. The spectrum is flux calibrated and is
  normalized at 5500 \AA. Prominent spectral lines are marked.}
\end{figure}

%figure 4
\begin{figure}
\figurenum{4}  
\plotone{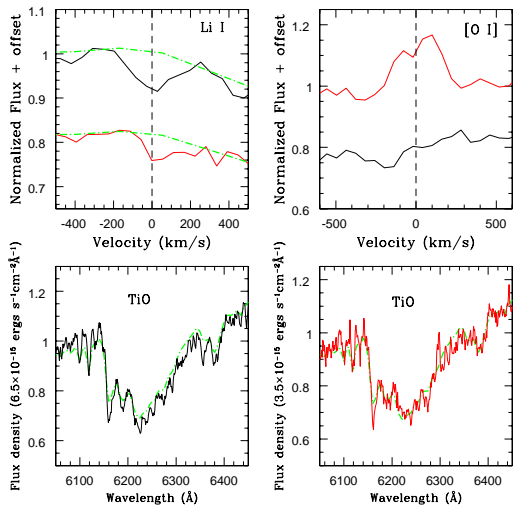}
\caption{\LiI~$\lambda$6708 and \OI~$\lambda$6300 line profiles of PDS 11A and PDS 11B (red) are
  shown in the upper panel. The lower panel shows the TiO $\lambda$6180 band of PDS 11A (left)
  and PDS 11B (right). Overplotted is the M2V template from Pickles stellar library in green.}
\end{figure}

We found \OI~$\lambda$6300 in emission in PDS 11B (Figure 4).
This line is often seen in the collimated jets driven by accreting
T Tauri stars \citep[e.g.][]{Hartigan95,Hartmann09}.
\citet{Whitelock95} also noticed \OI~$\lambda$6300 in the spectrum of PDS 11B. This emission line is
not seen in the spectrum of PDS 11A, both in our observations and that of \citet{Whitelock95}. 
\citet{Zuckerman14} have suggested that \OI~$\lambda$6300 is formed by the
photodissociation of OH molecules in the disk by far-ultraviolet stellar
photons. We find evidence for the presence of \HeI~$\lambda$4471 and $\lambda$5876 in emission in the
spectra of PDS 11B (Figure 3). \HeI~$\lambda$4471 is red shifted by 3 \AA~,
as seen from the spectra observed on 2016 February 13. This feature was absent during the observation
on March 15. \HeI~$\lambda$5876 is affected by molecular absorption bands. However, when
compared with the spectrum of PDS 11A, where \HeI~$\lambda$5876 is equally
affected by molecular band absorption, an emission component is seen at $\lambda$5876. It may be
noted that \citet{Whitelock95} identified \HeI~$\lambda$5876
emission in PDS 11B. It is possible that \HeI~emission lines
are formed from photoionization and subsequent
recombination in the accretion shock region, close to the stellar surface \citep{Zuckerman14}.
Hence, the presence of \OI~$\lambda$6300 and \HeI~emission lines in the spectrum of PDS 11B suggest that
it belongs to the class of accreting T Tauri stars. 

%figure 5
\begin{figure*}
\figurenum{5}
\plottwo{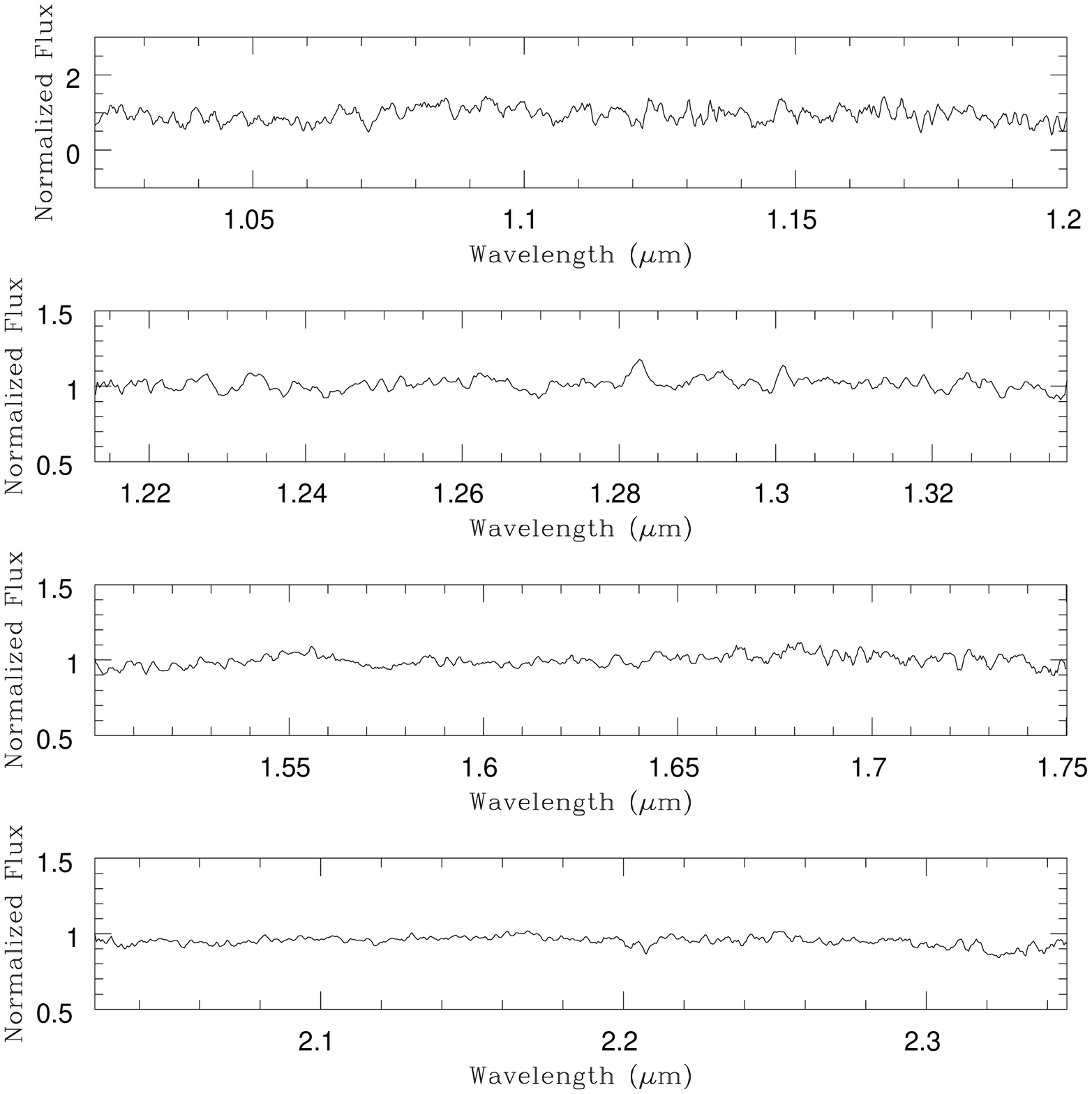}{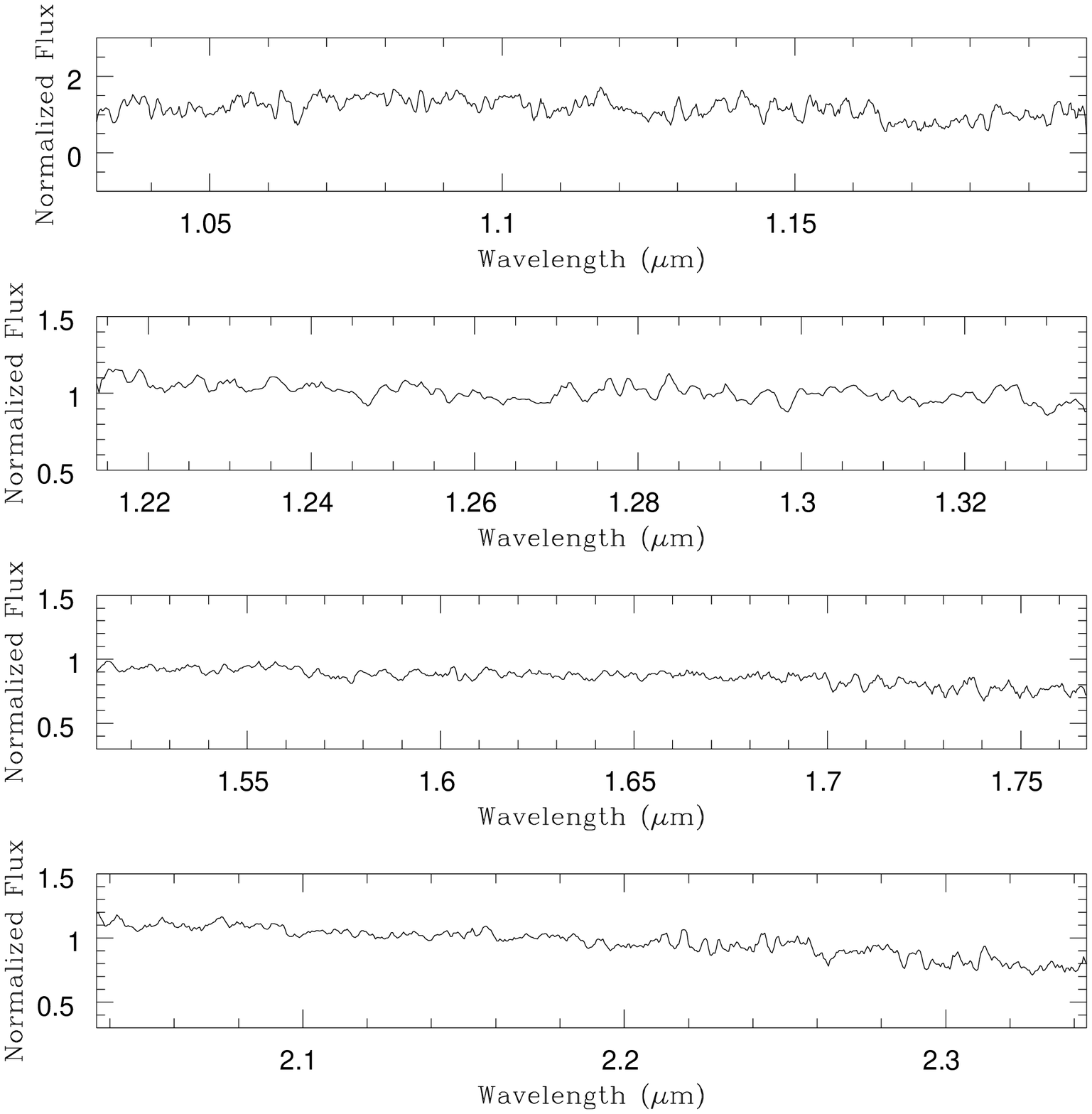}
\caption{Near-Infrared $Y$, $J$, $H$, $K$, spectra of PDS 11A (left) and PDS 11B (right)}
\end{figure*}

The flux calibrated $Y$, $J$, $H$, $K$ spectra of PDS 11A and PDS 11B,
normalized to band center values are shown in Figure 5. Since the signal
to noise is low, the spectra is smoothed to 5 points for display purpose. 
Pa$\beta$ is the only prominent spectral line found in PDS 11A, which is
present in emission with an EW of 3 \AA. 
No prominent emission or absorption features are present in PDS 11B. 

\subsection{Spectral type estimation}

We estimated the spectral type of PDS 11A and PDS 11B using TiO bands, the dominant
molecular absorption band in M-type stars. The spectral type has been
estimated from the TiO5 spectral index, which is
defined as the ratio of mean flux value in 7126$-$7135 \AA~region to that
at 7042$-$7046 \AA~wavelength region \citep{Reid95}. \citet{Reid95} 
derived a relation between the spectral type (SpT) and TiO5 index, 
SpT = $-$10.775$\times$TiO5 + 8.2. They suggested using 
this relation as a reliable spectral type estimator in the range K7 to M6.5
dwarfs. We have measured the flux values from the flux calibrated spectra obtained on
2016 February 13 and found TiO5 to be 0.60$\pm$0.01 for PDS 11A and 0.56$\pm$0.01 for PDS 11B. The
spectral type is estimated from the above mentioned relation whereby PDS 11A
is found to be M1.7$\pm$0.1 and PDS 11B to be M2.1$\pm$0.1. Since fractional subtypes
are not used for analysis, we will be approximating the spectral type of both stars as M2,
but suggest that PDS 11A can be earlier than PDS 11B by a fractional subtype.
Further, we compared the absorption strength of TiO $\lambda$6180 absorption feature with
the M2V spectral template from Pickles library (see Figure 4). A close match between the
TiO band strength of the object and the template spectrum support the M2 classification of PDS 11A and PDS 11B.
The early M-type classification is further supported by 
the absence of TiO $\lambda$8465 band in our spectra. This
band starts appearing in the spectra from M3 onwards and is used as a criteria
to classify objects with spectral type M3$-$L3 \citep{Slesnick06}. 

%table 2
\begin{deluxetable*}{llcccc}
\tablecaption{EW and FWHM of the major spectral lines in PDS 11A and PDS 11B. EW of emission lines are shown in negative \label{tab:table2}}
\tablehead{
\colhead{Spectral line} & \colhead{Date of obs.} & \twocolhead{PDS 11A} & \twocolhead{PDS 11B} \\
\colhead{}              & \colhead{}             & \colhead{EW (\AA)}   & \colhead{FWHM (\AA)} & \colhead{EW (\AA)} & \colhead{FWHM (\AA)}
}
\colnumbers
\startdata
Ca{\sc ii} K & 2016 Feb. 13 & -23$\pm$0.3 & 7.9$\pm$0.1 & -48$\pm$4 & 8.6$\pm$0.6\\
          & 2016 Mar. 15 & -28$\pm$3 & 9.5$\pm$1 & -38$\pm$6 & 7.5$\pm$2\\
Ca{\sc ii} H + H$\epsilon$ & 2016 Feb. 13 & -22$\pm$1 & 7.9$\pm$0.4 & -23$\pm$3 & 7.8$\pm$0.2\\
          & 2016 Mar. 15 & -17$\pm$2 & 7.6$\pm$0.3 & -24$\pm$4 & 8.0$\pm$1\\
H$\delta$ & 2016 Feb. 13 & -14.6$\pm$0.6 & 9.0$\pm$0.3 & -9.4$\pm$0.8 & 9.0$\pm$0.4 \\
          & 2016 Mar. 15 & -9.1$\pm$0.6 & 8.4$\pm$0.6 & -10$\pm$3 & 8.0$\pm$2 \\ 
H$\gamma$ & 2016 Feb. 13 & -12.8$\pm$0.5 & 8.9$\pm$0.2 & -10.8$\pm$0.6 & 9.3$\pm$0.2 \\
         & 2016 Mar. 15 & -11.4$\pm$0.4 & 9.0$\pm$1 & -11$\pm$2 & 8.5$\pm$1 \\ 
4474 (HeI?) & 2016 Feb. 13 & . & . & -2.4$\pm$0.1 & 6.2$\pm$0.3\\
          & 2016 Mar. 15 & . & . & . & . \\
H$\beta$ & 2016 Feb. 13 & -10.8$\pm$0.8 & 8.8$\pm$0.4 & -13.0$\pm$0.7 & 8.6$\pm$0.4 \\
         & 2016 Mar. 15 & -7.8$\pm$0.6 & 8.3$\pm$0.3 & -10$\pm$1 & 8.4$\pm$0.4\\
He{\sc i} 5876 & 2016 Feb. 13 &  . & . & -1.9$\pm$0.1 & 9.0$\pm$0.1 \\
          & 2016 Mar. 15 & . & . & -1.4$\pm$0.4 & 8.2$\pm$0.6 \\
Na{\sc i} (5890+5896) & 2016 Feb. 13 & 5.2$\pm$0.2 & 12.4$\pm$0.3 & 3.4$\pm$0.2 & 12.7$\pm$0.4\\
          & 2016 Mar. 15 & 6.0$\pm$0.2 & 13.5$\pm$0.5 & 3.2$\pm$0.3 & 13.8$\pm$0.5\\ 
\OI 6300 & 2016 Feb. 13 & . & . & -1.2$\pm$0.1 & 6.5$\pm$0.1 \\
                & 2016 Mar. 15 & . & . & -1.7$\pm$0.3 & 5.9$\pm$0.4 \\
H$\alpha$ & 2016 Feb. 13 & -23.0$\pm$0.8 & 7.6$\pm$0.1 & -27.0$\pm$0.7 & 7.9$\pm$0.2 \\
          & 2016 Mar. 15 & -25.0$\pm$0.5 & 7.7$\pm$0.2 & -25.0$\pm$0.5 & 8.3$\pm$0.2\\ 
Li{\sc i} 6708 & 2016 Feb. 13 & 0.45$\pm$0.05 & 5.2$\pm$0.1 & . & . \\
               & 2016 Mar. 15 & 0.40$\pm$0.04 & 5.0$\pm$0.3 & . & . \\
Na{\sc i} 8183 & 2016 Feb. 13 & 0.8$\pm$0.1 & 6.8$\pm$0.2 & 0.8$\pm$0.1 & 6.3$\pm$0.1 \\
               & 2016 Mar. 15 & 1.0$\pm$0.1 & 7.9$\pm$0.2 & 0.9$\pm$0.2 & 7.3$\pm$0.2\\
Na{\sc i} 8195 & 2016 Feb. 13 & 1.0$\pm$0.1 & 6.4$\pm$0.2 & 1.0$\pm$0.1 & 6.8$\pm$0.1 \\
               & 2016 Mar. 15 & 1.3$\pm$0.1 & 7.1$\pm$0.1 & 1.1$\pm$0.2 & 7.2$\pm$0.2\\
\enddata
\end{deluxetable*}

\subsection{H$\alpha$: Accretion indicator}

Historically, the strength of the H$\alpha$ line has been used to
distinguish between accreting classical T Tauri stars (CTTS) from weak-lined T Tauri stars (WTTS), where
the H$\alpha$ emission is due to chromospheric activity. An H$\alpha$ equivalent
width EW(H$\alpha$) $\sim$ 10 \AA~was set as the discrimination
boundary. However, because of the `contrast effect' of
the photosphere, no unique EW(H$\alpha$) value distinguishes all
CTTSs from WTTSs, and several authors have proposed 
EW(H$\alpha$) dividing line as function of spectral type
\citep{Martin98,White03,Barrado03}. \citet{White03} prescribed empirically determined
maximum EW(H$\alpha$) values observed for non-accreting
T Tauri stars for different spectral type ranges. 
\citet{Barrado03} proposed EW(H$\alpha$) values as a
function of spectral type derived from the observed saturation
limit for the chromospheric activity at Log(L$_{H\alpha}$/L$_{bol}$) = $-$3.3.
Figure 6 shows both \citet{White03} and \citet{Barrado03} criteria to distinguish
between CTTSs from WTTSs. Also shown in the figure are
the EW(H$\alpha$) values of PDS 11A and PDS 11B, which are well
above the lines depicting \citet{White03} and \citet{Barrado03} criteria. Thus, both PDS 11A
and PDS 11B are accreting and are CTTSs.

%figure 6
\begin{figure}
\figurenum{6}  
\plotone{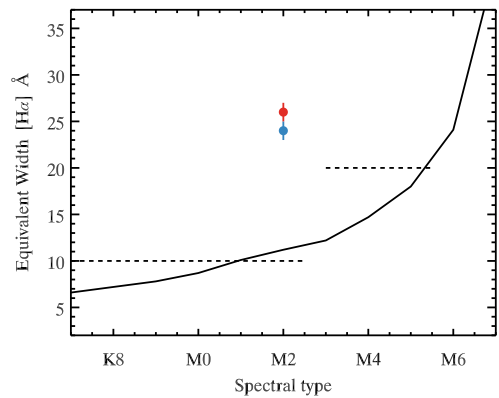}
\caption{The EW(H$\alpha$) criterion which distinguishes CTTSs from
WTTSs as a function of spectral type from 
\citet{White03} (dashed lines) and \citet{Barrado03} (solid
line) is shown. Observed EW(H$\alpha$) values of PDS 11A (blue solid circle) and
PDS 11B (red solid circle) from 2016 February 13 are also shown.}
\end{figure}

We computed the accretion rates for PDS 11A from the 
H$\alpha$ and H$\beta$ line luminosities using the 
empirical relations given by \citet{Herczeg08}, \citet{Fang09},
and \citet{Ingleby13}.
For the distance range listed in Table 3 (114$-$131 pc)
corresponding to the age range of 10$-$15 Myr, the accretion rates obtained
are in the range of 4.2$\times$10$^{-11}$ $-$ 5.0$\times$10$^{-10}$ M$_\odot$ yr$^{-1}$
with a median value of $\sim$~1.0$\times$10$^{-10}$ M$_\odot$ yr$^{-1}$.
The accretion rates we obtain for PDS 11A is significantly 
lower than that found for $<$~3 Myr old CTTSs of spectral type
M2 \citep[e.g.,][]{Ingleby13,Kim16}.
They are, however, quite similar to the accretion rates found for M2
members of the 10$-$15 Myr old moving groups \citep[e.g.,][]{Zuckerman14}.
The H{\sc i} line luminosity and the $L_{acc}$ estimated from 
them for the PDS 11A are significantly higher than that expected 
from chromospheric activity. Following \citet{Manara13}, the noise 
introduced in the estimated $L_{acc}$ due to chromospheric contamination
\citep[see eqn. 2 in][]{Manara13} is $<$~1.8$\times$10$^{-4}$~$L_\odot$, 
while the accretion luminosity of PDS 11A is
$\sim$~1.4$\times$10$^{-3}$~$L_\odot$, which is $\sim$~8 times higher, 
indicating that PDS 11A is accreting material from the disk.

From the accretion rates of PDS 11A,
we further estimated the expected line flux and equivalent widths 
for Pa$\beta$ and Br$\gamma$ lines using the 
empirical relations from \citet{Muzerolle98}, \citet{Calvet04} and \citet{Natta06}. 
For PDS 11A, the expected Pa$\beta$ EW is $\sim$2 \AA~(line flux $\sim$2.1$\times$10$^{-21}$ W cm$^{-2}$) 
and the expected Br$\gamma$ EW is $\sim$0.3 \AA~(line flux $\sim$3.1$\times$10$^{-22}$ W cm$^{-2}$).
From the observed spectra discussed in Sect. 3.1 we found that Pa$\beta$ EW is
around 3 \AA~whereas no clear emission is present in Br$\gamma$, which
agrees with these estimates. Thus the strength of the observed
\HI~lines in the optical and near-IR spectra of PDS 11A is consistent
with the star accreting at a rate of $\sim$10$^{-10}$ M$_\odot$ yr$^{-1}$.

\subsection{Age estimation of PDS 11A from \LiI~$\lambda$6708 EW}

The \LiI~$\lambda$6708 absorption feature is an indicator of youth and is often used as one of
the criteria to classify the source as a T Tauri star \citep{Bodenheimer65,Hamann92,Sergison13}.
We found evidence for the presence of \LiI~$\lambda$6708 line in the
spectrum of PDS 11A, with an EW of 0.43 \AA~(mean value of both epochs, Table 2).
In Figure 4, the presence of \LiI~$\lambda$6708 absorption is particularly evident from the
  comparison of line profile with that of M2V spectral template from Pickles library.
  It may be noted that a tentative detection of \LiI~$\lambda$6708 in PDS 11B is seen in Figure 4,
  when compared to the template spectrum, but can only be confirmed from spectra with better resolution
  and signal-to-noise. Henceforth we will be considering \LiI~$\lambda$6708 absorption only in PDS 11A.
  \citet{GregorioHetem92} also have reported \LiI~$\lambda$6708 
absorption line in the spectrum of PDS 11A with an EW of 0.71 \AA. The
identification of \LiI~$\lambda$6708 in the spectra of PDS 11A supports
its T Tauri membership.  

We have used lithium depletion boundary (LDB) technique to estimate the age of PDS 11A.
This method is model independent when compared to the age
estimation from stellar evolutionary models and is often used for precise
age estimation of young stars in moving groups \citep{Soderblom10,Song02,Binks14}. 
From the measured \LiI~$\lambda$6708 EW it is possible to estimate the
age of the T Tauri star using LDB technique. Since we know the spectral type
of our object of interest, it is possible to set an age limit during
which \LiI~$\lambda$6708 absorption line is present in the
spectra. The stellar models of \citet{Baraffe15} show that the 
surface lithium is depleted in M2 stars such as PDS 11A
\citep[whose effective temperature (T$_{eff}$) is 3490 K;][]{Pecaut13}
at an age of 15 Myr. This means that since \LiI~$\lambda$6708 is present
in PDS 11A, the age should be less than 15 Myr.

To see whether the age of PDS 11A matches with young stars in moving groups, we analyzed the
\LiI~$\lambda$6708 EW measurements of all the known members of the nine nearby moving
groups identified so far, from \citet{daSilva09}. We have used those
values along with the stellar effective temperature
\citep[as given in][]{daSilva09} to analyze the \LiI~$\lambda$6708 EW variation with
temperature (Figure 7). In Figure 7, PDS 11A is located between 10 Myr old TW
Hya association and 24 Myr old $\beta$~Pic moving group.
It may be noted there are no moving groups with \LiI~EW measurements of stars between 10 and 24 Myr.
Hence, from LDB method and the analysis of \LiI~EW distribution in moving groups,
we found that PDS 11A has an age of 10$-$15 Myr, which will be used in further discussion.

%figure 7
\begin{figure}
\figurenum{7}
\plotone{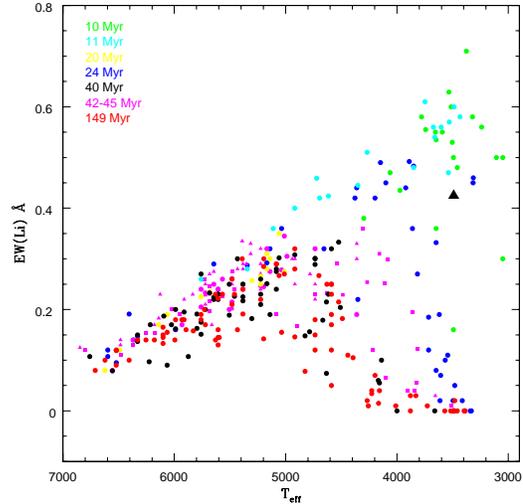}
\caption{The Li$\lambda$ 6708 EW of members of various moving groups
  is shown as a function of their effective temperature
  \citep[taken from][]{daSilva09}. PDS 11A is shown as black triangle. 
The members of various groups are color coded based on their ages; 
TW Hya (10 Myr, green), $\beta$Pic (24 Myr, blue), Tucana (45 Myr, magenta), 
Columba (42 Myr, magenta), Carina (45 Myr, magenta), $\epsilon$Cha (11 Myr, cyan),
Octans (20 Myr, yellow), Argus (40 Myr, black), AB Dor (149 Myr, red). The ages of each of the
moving groups is taken from the recent compilation by \citet{Bell15}. The
age of Argus and Octans association is not listed in \citet{Bell15} and hence we used
the age given in \citet{daSilva09}.}
\end{figure}

PDS 11 is not the first case of a T Tauri binary system where 
only one of the components show \LiI~$\lambda$6708 absorption. 
\citet{Song02} found that \LiI~$\lambda$6708 absorption feature is
seen in the secondary of the HIP 112312 (GJ 871.1 A and B) pre-MS binary system whereas it is
absent in the primary. They have used this system, which is about 12 Myr old
and at a distance of 24 pc, to study the LDB in pre-main sequence stars. 
The spectral type of GJ 871.1 A and B (M4 and
M4.5) is closer to that of PDS 11A and PDS 11B (M2).
It is intriguing to see why \LiI~$\lambda$6708 is present in absorption
only in one component of the binary system, if they are of similar age. 
We note that presence of \LiI~$\lambda$6708 
in absorption is not a necessary condition to
classify the object as a T Tauri star. \citet{Baraffe10} have cautioned that lithium
may not always be a reliable age indicator since the 
lithium abundance depends on the accretion history of the star. 
Episodic accretion in young stars can increase the central temperature due
to which lithium can get severely depleted \citep{Chabrier96}. 

\subsection{Stellar parameters}

From our photometry we estimated $V$ and $(B-V)$ values of PDS 11A and PDS
11B as 14.75$\pm$0.03 (m$_{V1}$), 1.43$\pm$0.05, and 14.98$\pm$0.03 (m$_{V2}$),
1.28$\pm$0.05, respectively.
The intrinsic $(B-V)$ color of PDS 11A and PDS 11B is 1.46,
considering that both are M2 stars \citep{Pecaut13}.
Our observed $(B-V)$ colors are found to be bluer by
0.03 mag and 0.18 mag than
the intrinsic values, for PDS 11A and PDS 11B, respectively. 
This has been noticed in previous studies and could be caused by
the lower gravity of pre-MS stars with respect to the
dwarfs \citep{Song02}. From the observed $(B-V)$ values, 
the color excess $E(B-V)$ of both the stars is found to be negative and hence will be
considered as zero from now on. This is understandable since these are high
Galactic latitude objects and hence suffer little extinction.

Comparison with 10$-$15 Myr isochrones from \citet{Baraffe15}
for an M2 star (T$_{eff}$ = 3490 K) indicates a mass of 0.4 M$_\odot$ and
luminosity in the range 0.089$-$0.067 L$_\odot$ (log L/L$_\odot$ = $-$1.05 -- $-$1.17) for PDS 11A.
These luminosity values imply bolometric magnitude M$_{bol}$ in the range of 7.37$-$7.67,
from which absolute $V$ magnitude (M$_{V1}$) is obtained using the bolometric correction
of -1.80 for M2 stars given in \citet{Pecaut13}.
The observed V magnitude, m$_{V1}$ = 14.75, and M$_{V1}$ indicate a distance of 114$-$131 pc for PDS 11A.
The estimated stellar parameters are given in Table 3. Similar estimates are not possible for PDS 11B as
\LiI~$\lambda$6708 is not present in the spectra.

%table 3
\begin{deluxetable}{lcc}
\tablecaption{Stellar properties \label{tab:table3}}
\tablehead{
\colhead{Reference} & \colhead{PDS 11A} & \colhead{PDS 11B}
}  
\colnumbers
\startdata
Sp.type & M1.7$\pm$0.1 & M2.1$\pm$0.1 \\
T$_{eff}$ (K) & 3490 & 3490 \\
L$_{bol}$ (L$_\odot$) & 0.067$-$0.089 & . \\
$E(B-V)$ & 0 & 0 \\
Distance (pc) &  114$-$131 & . \\
Age (Myr) & 10$-$15 & . \\ 
Mass (M$_\odot$) & 0.4 & . \\
V$_r$ (\kms) & 21$\pm$3$^1$ & 10$\pm$8$^2$ \\
$\mu_\alpha$$^3$ (mas~yr$^{-1}$) & 6.0$\pm$4.3 & 5.7$\pm$3.4 \\ 
$\mu_\delta$$^3$ (mas~yr$^{-1}$) & -1.4$\pm$4.8 & 4.6$\pm$2.5 \\
\enddata
\tablecomments{References: 1 -- \citet{GregorioHetem92}, 2 -- \citet{Whitelock95}, 3 -- \citet{Qi15}}
\end{deluxetable}

%figure 8
\begin{figure}
\figurenum{8}  
\plotone{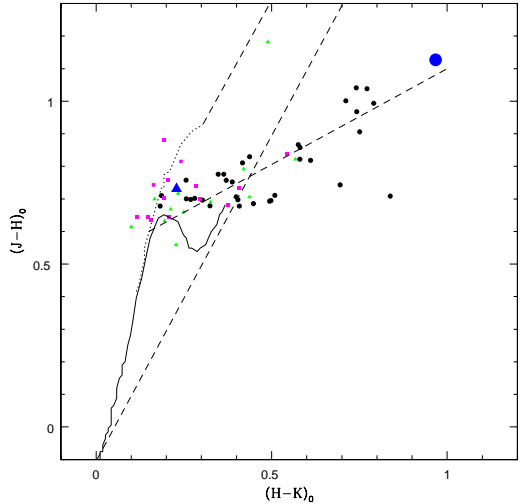}
\caption{2MASS $(J-H)_0$ vs $(H-K)_0$ color-color diagram: PDS 11A 
and PDS 11B are shown in open blue triangle and circle, respectively. The
sample of CTTS are shown in black circles, WTTS in magenta squares and TD
candidates in green diamonds.
Main sequence and giant sequence, shown in solid and dotted lines, respectively,
is from \citet{Koornneef83}, which is converted to 2MASS system using the
transformation relations from \citet{Carpenter01}. The CTTS location is
from \citet{Meyer97} and is shown in dot-long dash line. The $(J-H)$, $(H-K)$ colors
are reddening corrected using the relation from \citet{Rieke85}.}
\end{figure}

\subsection{Infrared excess}

Infrared excess in the energy distribution is one of the defining criteria to
identify T Tauri stars among the sample of low mass stars \citep{Calvet98,Meyer97}. We made use of the
archival 2MASS data to estimate the near-IR $(J-H)$, $(H-K)$ colors and use
them to assess near-IR excess in PDS 11A and PDS 11B.
Figure 8 shows the location of PDS 11A \& 11B in the $(J-H)$$-$$(H-K)$ color-color
diagram. For assessing the nature of near-IR excess in PDS 11 components, we
also show CTTS, WTTS and transitional disk candidates in the figure. The 32
CTTS shown are of spectral types M0$-$M3 from the Taurus star forming region,
identified from \citet{Furlan06} and 
\citet{Furlan11}. Also, included are 14 WTTS in the spectral type range
M0$-$M3 from the Taurus and Chamaeleon star forming 
regions \citep{Furlan11,Manoj11}. We have used the 2MASS colors of a 
sample of 16 TD candidates in Taurus and Chamaeleon I, taken 
from \citet{Kim13}. 
It is immediately evident from Figure 8 that PDS 11B shows considerable IR
excess and is found to be on the CTTS locus.
The $(J-H)_0$, $(H-K)_0$ colors of the object seems to be higher than the sample
of CTTS used for this analysis. However, since the location do not contain T
Tauri stars other than CTTS, it is pretty clear that PDS 11B belong to CTTS
category. The near-IR excess of PDS 11A is similar to the WTTS/TDs, suggesting
significantly less hot dust material around it.
In order to see whether our stars have any analogs in any of the moving group
in terms of near-IR colors, we have represented them in $(J-H)$ versus
$(H-K)$ color-color diagram. We have included stars in the spectral range M0$-$M3, from known moving groups listed in 
\citet{Zuckerman04} and \citet{Torres08}. As seen from Figure 9, almost all the members of
various moving groups are found to be clustered near the main sequence, similar to PDS 11A.
The extreme type of IR excess seen in PDS 11B is generally not seen in any other association members. In summary, both PDS 11A and PDS 11B are found to lie on the CTTS locus, indicating the presence of warm circumstellar dust around them.  In addition to the near-IR excess, both the stars are accreting (Sect. 3.3) and show veiling in the observed spectra (discussed in Sect. 4.1), qualifying them as classical T Tauri stars.  PDS 11B show very high near-IR excess and lies at the extreme end of the CTTS locus. Since none of the known CTTS display such high near-IR excess (see Figure 8), it is worth exploring the nature of PDS 11B from further studies.

%figure 9
\begin{figure}
\figurenum{9}  
\plotone{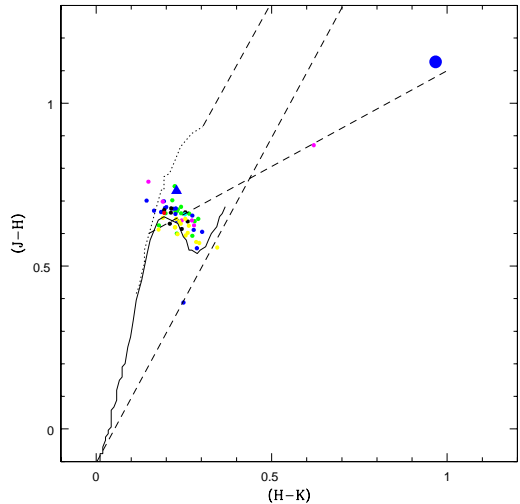}
\caption{2MASS $(J-H)$ vs $(H-K)$ color-color diagram: PDS 11A
  and PDS 11B are shown in open blue triangle and circle, respectively. 
The sample of stars from moving groups are shown in different colored filled circles; 
TW Hya (green), $\beta$~Pic (blue), Tucana (black),
Columba (cyan), $\epsilon$~Cha (magenta), Argus (red), AB Dor (yellow).
$(J-H)$ and $(H-K)$ colors are not dereddened since color excess values
of moving groups are not available. All the other sequences are same as in Figure 8.}
\end{figure}

\subsection{Spectral Energy Distribution}

The Spectral Energy Distribution (SED) of PDS 11A and PDS 11B is constructed
with the available photometric data given in Table 4. The SEDs are shown in
Figures 10. We have used BT-Settl
model atmospheres corresponding to the temperature (T$_{eff}$) and 
gravity (log $g$) values of of PDS 11A and PDS 11B. Even though 
T$_{eff}$ of both the stars are 3490 K, we have taken the BT-Settl
atmosphere for 3500 K, which is the closest temperature for which stellar
atmosphere is available.
Generally, in the case of pre-MS stars, model atmospheres corresponding to log($g$) = 4.5
is used for SED analysis. We have verified this in the case of PDS 11A,
which has a log($g$) value of 4.34 from the stellar models of \citet{Baraffe15}. Since BT-Settl
model atmospheres corresponding to T$_{eff}$ = 3490 K and log($g$) = 4.34 is unavailable,
we have used the nearest combination of T$_{eff}$ = 3500 K and log($g$) = 4.5 for SED analysis. 

%figure 10
\begin{figure}
\figurenum{10}  
\plotone{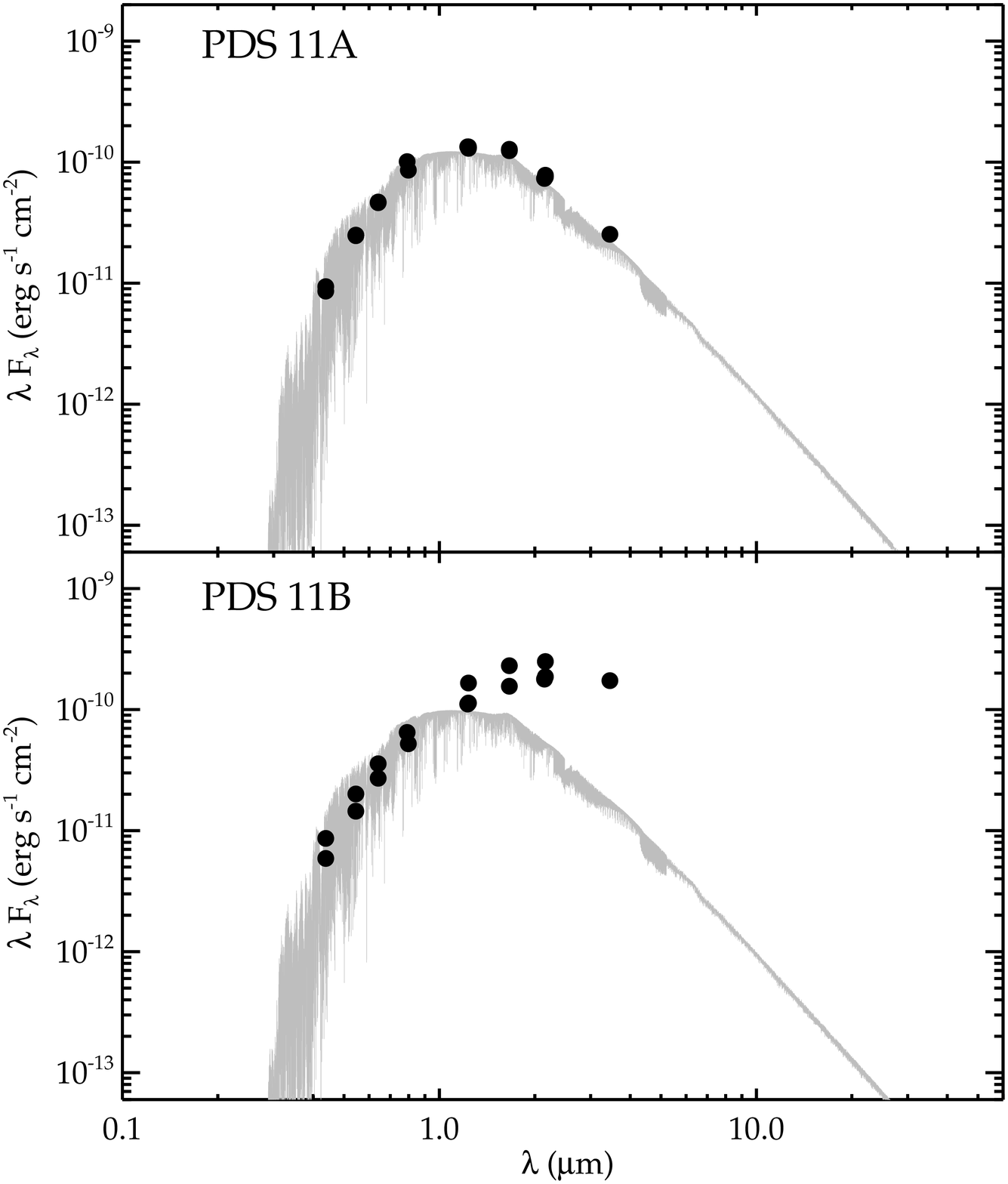}
\caption{Observed SEDs of PDS 11A and PDS 11B. Photometry at all the epochs are plotted.
  For PDS 11A they fall on top of each other, while PDS 11B show significant variability.
  Also shown are the BT-Settl model for T$_{eff}$ = 3500 K and log(g) = 4.5.
  The model is normalized to the observed J-band flux.}
\end{figure}

PDS 11A do not show much of IR excess whereas PDS 11B shows considerable excess
with the SED rising in near-IR itself (Figure 10).
We found that mid-infrared magnitudes are available from
$WISE$ mission for PDS 11B, but the beam size includes PDS 11A as
well. Comparison of $J$, $H$, $K$ magnitudes of both stars from
2MASS and \citet{Whitelock95} indicate that most 
of the excess emission is coming from PDS 11B. PDS 11B is about
2.2 mag brighter than PDS 11A in $L$ band \citep{Whitelock95}.
PDS 11B shows considerably high $(J-H)$ and $(H-K)$ color excess ($\sim$~1
mag), which is even higher than that for CTTS in star forming region (Figure 8).
We note that in addition to disk excess,
the IR excess in PDS 11B can also be contributed by an unseen late-type companion,
which demands further investigation.   

%table 4
\begin{deluxetable}{lccc}
\tablecaption{Available data \label{tab:table4}}
\tablehead{
\colhead{Reference} & \colhead{Band} & \colhead{PDS 11A} & \colhead{PDS 11B} 
}
\colnumbers
\startdata
This work & $B$ & 16.18$\pm$0.04 & 16.27$\pm$0.04 \\
          & $V$ & 14.75$\pm$0.03 & 14.98$\pm$0.03 \\
          & $R$ & 13.69$\pm$0.03 & 13.98$\pm$0.03 \\
G92       & $B$ & 16.27 & 16.68 \\
          & $V$ & 14.76 & 15.34 \\
          & $R$ & 13.70 & 14.28 \\
          & $I$ & 12.55 & 13.09 \\ 
DENIS     & $I$ & 12.433$\pm$0.03 & 12.915$\pm$0.03 \\
          & $J$ & 11.159$\pm$0.06 & 11.361$\pm$0.06 \\
          & $K$ & 10.259$\pm$0.06 &  9.293$\pm$0.06 \\
2MASS     & $J$ & 11.182$\pm$0.026 & 11.332$\pm$0.029 \\	
          & $H$ & 10.451$\pm$0.026 & 10.205$\pm$0.029 \\	
          & $K$ & 10.222$\pm$0.019 &  9.238$\pm$0.021 \\
W95       & $J$ & 11.152$\pm$0.036 & 10.919$\pm$0.189 \\
          & $H$ & 10.422$\pm$0.047 & 9.781$\pm$0.174 \\
          & $K$ & 10.185$\pm$0.039 & 8.923$\pm$0.173 \\
          & $L$ &  9.960$\pm$0.061 & 7.872$\pm$0.125 \\
\enddata
\tablecomments{References: DENIS -- DENIS Consortium 2005, 2MASS -- \citet{Cutri03},
  G92 -- \citet{GregorioHetem92}, W95 -- \citet{Whitelock95}; L mag given in SAAO system.}
\end{deluxetable}

\section{Discussion}

Our analysis so far suggest that both components of the PDS 11 system are of
similar spectral type, M2. Both PDS 11A and PDS 11B show strong H$\alpha$ emission,
confirming their CTTS status.
While PDS 11B show strong excess emission in the near-IR, PDS 11A show
no or weak excess, indicating the that no hot dust is present close to the
star. Intriguingly, PDS 11A show lithium absorption, suggesting that it is
10$-$15 Myr old; PDS 11B do not show lithium absorption.
Below we discuss more aspects on veiling and binarity of this system. 

\subsection{Veiling estimate from \CaI~$\lambda$4227}

The presence of excess continuum emission, referred to as veiling, 
is often observed in classical T Tauri stars \citep{Joy49,Johns-Krull01}. 
In the magnetospheric accretion model, veiling is due to the dissipation of energy in the
post-shock region at the base of the magnetic funnel, close to the stellar
surface \citep{Koenigl91,Hartmann94,Calvet98}.   
\citet{Herczeg14} suggested a method to estimate veiling in T Tauri stars. 
They found that measured EW
values of Ca{\sc i} $\lambda$4227 absorption line follow a relation with
spectral type, of the form EW(\CaI) = $-$189.218 + 7.36$x$ $-$ 0.072$x^2$. 
Since both PDS 11A and PDS 11B are of M2 spectral type, `$x$' corresponds to
60 \citep{Herczeg14}. Hence the expected \CaI~$\lambda$4227 EW should be
6.8 \AA, if veiling is not present. In the case of PDS 11A, 
EW(\CaI) measured on 2016 February 13 agrees with the expected value,
whereas it changed to 5.6$\pm$0.2 \AA, when measured from the spectra taken
on 2016 March 15. During the first epoch, the measured EW value of 
\CaI~$\lambda$4227 is almost similar to that expected if veiling is not
present. But in the second epoch, the line gets veiled, suggesting that
emission from accretion continuum is filling-in the absorption line.  
The situation is even more interesting in the case of PDS 11B
which show EW(\CaI) value of 6.0$\pm$0.4 from the spectra taken on
2016 February 13 and the line is almost not visible in the spectra taken on 2016
March 15. Since the measured EW of \CaI~$\lambda$4227 is considerably
lower than the empirical estimates, PDS 11B must be undergoing strong
veiling. We estimated the effect of veiling on spectral type determination 
  for both the stars using the EW(\CaI) relation given in \citet{Herczeg14}.
  We found that the spectral type of PDS 11A could change from M2 to M1 for a change in
  EW(\CaI) from 6.8 to 5.6 \AA.
  In the case of PDS 11B, the \CaI~$\lambda$4227 line is almost completely veiled and hence
  the spectral type can shift to earlier type by $\sim$5 subtypes.
  
The fact that the spectra of both the stars in PDS 11 are veiled has
important implications. 
Firstly, this supports our result that PDS 11A and PDS 11B are CTTS in
active accretion phase. Secondly, this explains the absence of \LiI~$\lambda$6708
absorption line in PDS 11B. It is quite possible that PDS 11B is of the same age
as PDS 11A and \LiI~$\lambda$6708 is present, but the line is filled in because of strong
veiling. We examined this aspect by comparing the lithium abundance values listed
  in \citet{Baraffe15} with the shift in spectral type due to veiling. In the previous paragraph,
  we described that veiling was present in the spectrum of PDS 11B obtained on 2016 February 13, while
  PDS 11A did not show evidence for veiling during that epoch. The equivalent width of \CaI~$\lambda$4227 in PDS 11B
  was found to be 6 \AA, whereas the expected value in the absence of veiling is 6.8 \AA.
  This reduction in absorption strength due to veiling will shift the
  spectral type from M2 to M1 for PDS 11B, according to the EW(\CaI) relation given in \citet{Herczeg14}.
  Thus, based on EW(\CaI), the spectral type of PDS 11B was of M1 and PDS 11A of M2 type, for the epoch 2016 February 13.
  The age of PDS 11A is estimated in the range 10$-$15 Myr. From \citet{Baraffe15} stellar models
  it is seen that at an age of 10 Myr, the ratio of surface lithium abundance to initial abundance
  is $\sim$27 times higher in an M2 star like PDS 11A, when compared to M1 star like PDS 11B. Since we found \LiI~$\lambda$6708
  absorption in PDS 11A to be 0.43 \AA, the EW(\LiI) expected for PDS 11B is around 0.02 \AA.
  This value is far below the detection limit of our instrument, which supports our proposition
  that \LiI~$\lambda$6708 might be present in PDS 11B, but not visible in the spectrum due to strong veiling.
 
\subsection{Possible association with a moving group}

We have used the Banyan II webtool\footnote{www.astro.umontreal.ca/~gagne/banyanII.php}
to check whether our candidates are
associated with any of the nearby kinematic groups. 
Banyan II is a Bayesian analysis tool which makes use of the position and 
space velocities of the object to assess the match with the database of nearby
($<$100 pc) moving groups, younger than 100 Myr \citep{Gagne14,Malo13}. 
The heliocentric radial velocity and proper motion of PDS 11A and PDS 11B are taken from
the literature and are given in Table 3. From the analysis we found that PDS 11A 
and PDS 11B are not associated with any known moving
group. Banyan II analysis gives 100\% probability that PDS 11A and PDS 11B
belong to young field population \citep{Gagne14}. It is quite possible that our object
parameters may not match with that of any known association and hence has been ascribed
to young field population.   

\subsection{PDS 11: a wide binary classical T Tauri system?}

Although the PDS 11 system has been treated as binary in the literature \citep{GregorioHetem92},
it is yet to be demonstrated that PDS 11A \& 11B are gravitationally bound.
It has been identified as visual binaries in
Washington Double Star catalog \citep{Mason01},
but that do not guarantee them being gravitationally bound.
Instead, we found that the proper motion in RA and Dec for both the stars
are similar within uncertainties (Table 3). This suggests that PDS 11A and PDS 11B form a binary
system. Considering PDS 11 at a distance of 114$-$131 pc, with a
separation of 8.8$\arcsec$ between the components \citep{Mason01}, the physical separation 
between PDS 11A and PDS 11B is $\sim$~1003$-$1153 AU. The proximity of two CTTS at such a separation
argues against chance alignment and suggests that they are part of a bound system.
Hence, the age and distance estimated for PDS 11A can as well be applied to PDS 11B.    

From a study of T Tauri stars in Taurus-Auriga star-forming region,
\citet{Bertout07} obtained a relation to estimate the lifetime of the disk in T Tauri stars ($\tau_d$), 
$\tau_d$ = 4$\times$10$^6$(M/M$_\odot$)$^{0.75}$, in terms of the mass of the parent
star (M). Since the mass of PDS 11A and PDS 11B are estimated to be 0.4 M$_\odot$,
the disk lifetime is around 2.0 Myr. 
Evidently it is quite puzzling how a disk which harbors enough gas and dust
survive in 10$-$15 Myr old system like PDS 11.
T Tauri binary systems with disks at ages older than the typical disk
dissipation timescales have been reported in the literature.
Most of them belong to nearby ($<$ 100 pc) young moving groups.
They are: 8 Myr old binary systems
TW Hya \citep{Teixeira08}, HR 4796 \citep{Kastner08}, TWA 30 \citep{Looper10a,Looper10b},
T Cha \citep{Kastner12}, 20 Myr old V4046 Sgr \citep{Kastner11}, and
LDS 5606 \citep{Rodriguez14}. These objects belong to class of wide
binaries, where the separation between the components is in the
range 1.7 kau \citep[for LDS 5606,][]{Rodriguez14}
to 41 kau \citep[for TW Hya$-$TWA 28 system,][]{Teixeira08}.
Among them, only TWA 30 and LDS 5606 are T Tauri binary systems in which both components are accreting.
The binary components of TWA 30 system are separated by $\sim$3400 au, show nearly edge-on orientation and
are of similar spectral type, M5 and M4, respectively \citep{Looper10b,Principe16}.
LDS 5606 is an M5/M5 T Tauri binary system at a separation of $\sim$1700 au and
they are members of $\beta$~Pic moving group \citep{Rodriguez14,Zuckerman14}. 

To summarize, PDS 11 is the third such system, after TWA 30 and LDS 5606, which belong
to the interesting class of old, dusty,
wide binary classical T Tauri systems in which both components undergo active accretion.
It is quite possible that PDS 11A is a transition disk candidate since it is accreting and
lacks hot dust material close to the star (see Sect. 3.3 \& 3.6). However, further studies are
needed to confirm other transition disk properties such as mid- and far-IR excess
and the presence of outer disk in PDS 11A. Also, further observations are needed to assess
  whether the near-IR excess in PDS 11B is entirely due to circumstellar material
  or due to the contribution from late-type companion. If confirmed, this would be the first known example of
a $>$10 Myr old binary system, where one of the components harbor
a radially continuous full disk, while the other is surrounded
by a disk with inner hole or gap. 

\section{Conclusion}

We have analyzed the star/disk properties and derived the spectral type
of the T Tauri binary system PDS 11 from optical photometry,
spectroscopy and infrared spectroscopic observations.
Our analysis indicates that PDS 11 is the new addition, after TWA 30 and LDS 5606,
to the interesting class of old, dusty, wide binary classical T Tauri systems
in which both components are actively accreting. The main
conclusions from this study are listed below. 

\begin{itemize}

\item{The spectral type of PDS 11A and PDS 11B were not known.
  We have classified both as M2-type making use of the TiO $\lambda$7050 band feature,
  with the aid of TiO5 index and the relations from \citet{Reid95}.}
  
\item{PDS 11A and PDS 11B are found to have H$\alpha$ emission
  strength of $\sim$25 \AA~, which is higher than the threshold value of
  chromospherically active stars. The median accretion rate derived from H$\alpha$ emission line is around
  $\sim$~1.0$\times$10$^{-10}$ M$_\odot$ yr$^{-1}$ for PDS 11A. PDS 11B show very high near- and mid-infrared excess.
  The emission lines of \OI~$\lambda$6300 and \HeI~$\lambda$5876, indicative of accretion process, are present
  in the spectrum of PDS 11B. All these evidences conclusively prove that PDS 11A and PDS 11B are classical T Tauri stars.
  It needs to be assessed from future studies how an active accretion disk sustains in a 10$-$15 Myr old system like PDS 11A.}
  
\item{We found that PDS 11A is less than 15 Myr from age dating using lithium depletion boundary method.
  Further, from the comparison of \LiI~$\lambda$6708 EW with that of young stars in moving groups, the age is constrained
  in the range 10$-$15 Myr.}  

\item{Since \LiI~$\lambda$6708 is not present in the spectrum of PDS 11B, the stellar parameters other
  than spectral type and temperature were not determined. However, since we prove that PDS 11A and
  PDS 11B form a binary system, age and distance of PDS 11B is taken to be similar to PDS 11A.}

\item{From our analysis PDS 11 is identified as a binary system with component masses of 0.4 M$_\odot$,
    luminosity of 0.067$-$0.089 L$_\odot$ and at a distance of 114$-$131 pc.}\\
  
\item{From the analysis with Banyan II webtool, 
PDS 11A and PDS 11B were not identified as members of any known moving group
and hence is considered as a young field binary system.}

\end{itemize}

\section*{Acknowledgments}

We would like to thank the staff at IAO, Hanle and its remote
control station at CREST, Hosakote for their help during the observation
runs. This research uses the SIMBAD astronomical data base service operated at
CDS, Strasbourg. This publication made use data of 2MASS, which
is a joint project of University of Massachusetts and the Infrared
Processing and Analysis Centre/California Institute of Technology,
funded by the National Aeronautics and Space Administration and
the National Science Foundation.

\end{document}